\documentclass[runningheads]{cl2emult}

\usepackage{makeidx}  % allows index generation
\usepackage{graphicx} % standard LaTeX graphics tool
\usepackage{subeqnar} % subnumbers individual equations
\usepackage{multicol} % used for the two-column index
\usepackage{cropmark} % cropmarks for pages without
\usepackage{physbb}   % centering for subeqnar.sty
\makeindex            % used for the subject index
\def\cross{\times}
\def\implies{\Rightarrow}
\def\Lemaitre{Lema\^\i{}tre}
\def\Painleve{Painlev\'e}
\def\Exercise{{\em Exercise:\ }}

\begin{document}
\title*{Acoustic Black Holes}
\toctitle{Acoustic Black Holes}
% allows explicit linebreak for the table of contents
%
%
\titlerunning{Acoustic Black Holes}
% allows abbreviation of title, if the full title is too long
% to fit in the running head
%
\author{Matt Visser}
\authorrunning{Matt Visser}
% if there are more than two authors,
% please abbreviate author list for running head
%
%
\institute{Washington University, Saint Louis, Missouri 63130-4899, USA}
\maketitle              % typesets the title of the contribution
%---------------------------------------------------------------------------
\begin{abstract}
Acoustic propagation in a moving fluid provides a conceptually clean
and powerful analogy for understanding black hole physics. As a
teaching tool, the analogy is useful for introducing students to both
General Relativity and fluid mechanics. As a research tool, the
analogy helps clarify what aspects of the physics are {\em kinematics}
and what aspects are {\em dynamics}. In particular, Hawking radiation
is a purely kinematical effect, whereas black hole entropy is
intrinsically dynamical.  Finally, I discuss the fact that with
present technology acoustic Hawking radiation is {\em almost}
experimentally testable.\\
{\em To appear in the Proceedings of the 1998 Peniscola Summer School on Particle 
Physics and Cosmology. (Springer-Verlag).}
\end{abstract}
%--------------------------------------------------------------------------

%-------------------------------------------------------------------------
\section{Developing the analogy}
%-------------------------------------------------------------------------
%
To ask how sound waves propagate in a moving fluid is a surprisingly
subtle question that rapidly introduces one to the full power and
complexity of curved-space Lorentzian differential geometry
\cite{Unruh81,Jacobson91,Jacobson93,Visser93a,Unruh94,Visser98a,Visser98b}.
A sound wave propagating in a flowing fluid shares many of the
properties of a minimally coupled massless scalar field propagating in
a non-flat (3+1)--dimensional Lorentzian geometry. This partial
isomorphism is the basis of a very useful analogy whereby {\em parts}
of General Relativity can be identified with {\em parts} of
non-relativistic fluid mechanics. Kinematic aspects of GR, such as the
existence of event horizons, carry over to fluid mechanics (event
horizons map into the boundaries of regions of supersonic
flow). Dynamic aspects of GR (the Einstein equations) do {\em not}
carry over. The analogy is not an identity, nevertheless enough
features are shared in common to make the model very useful, and
rather entertaining.  (Since this is a summer school, I will be very
pedagogical and will set out a number of exercises as we work through
the details.)

%------------------------------------------------------------------------
\subsection{Ingredients}
%-------------------------------------------------------------------------
%
The basic idea is to consider a non-relativistic, irrotational,
barotropic fluid.  The fluid should be irrotational since in this case
the velocity is completely specified by a scalar field, (which does not
have to be single-valued):
\begin{equation}
\nabla\cross\vec v = 0; \qquad \implies \qquad \vec v = \nabla \psi.
\end{equation}
Thus there is hope that the sound waves, which we shall soon see are
merely linearized fluctuations in the velocity field, can also be
described by a scalar field. There is of course nothing physically
wrong with non-scalar sound (vector sound?)  but the equations then
become so unmanageable as to be completely unwieldy. In the meantime,
as long as whatever vorticity happens to be present is confined to
thin vortex tubes, the present analysis is perfectly capable of
handling everything outside the vortex core.  The fluid should be
barotropic: The pressure should be a function of the density
only. This requirement makes sure that the pressure forces do not
generate vorticity---it guarantees that an initially irrotational
fluid will remain irrotational.  An additional simplifying assumption,
for the purposes of this talk, is that the viscosity is zero (inviscid
fluid). This is merely a technical simplification, and the
complications attendant on introducing viscosity into the system are
already understood~\cite{Visser98a}. (Viscosity acts as an explicit
breaking term for the acoustic Lorentz invariance, and acoustic
Lorentz invariance becomes an approximate symmetry that improves in
the low-frequency low-wavenumber limit.)  The relevant dynamical
equations are:\\ (1) the continuity equation
\begin{equation}
{\partial \rho \over\partial t} + \nabla\cdot(\rho \vec v) =0;
\end{equation}
(2) the (zero-viscosity) Euler equation
\begin{equation}
\rho \left( 
{\partial \vec v \over \partial t} + (\vec v \cdot \nabla) \vec v 
\right) = - \nabla p;
\end{equation}
(3) some barotropic equation of state
\begin{equation}
p = p(\rho).
\end{equation}
\Exercise (Easy) Check that this is a closed system of
equations. Check that if the vorticity is initially zero it will remain
so.

%-------------------------------------------------------------------------
\subsection{Manipulations}
%-------------------------------------------------------------------------
%
Now pick some arbitrary but exact solution [$\rho_0(t,x)$, $p_0(t,x)$,
$\psi_0(t,x)$] of the equations of motion. Treat this exact solution
as a background field and ask how linearized fluctuations around this
background behave. Write
\begin{eqnarray}
\rho(t,x) &=& \rho_0(t,x) + \epsilon \; \rho_1(t,x) + \cdots,
\\
p(t,x) &=& p_0(t,x) + \epsilon \; p_1(t,x) + \cdots,
\\
\psi(t,x) &=& \psi_0(t,x) + \epsilon \; \psi_1(t,x) + \cdots.
\end{eqnarray}
The equations of motion for these linearized fluctuations are
\begin{eqnarray}
{\partial \rho_1 \over \partial t} + 
\nabla \cdot \left( \rho_1 \nabla \psi_0 + \rho_0 \nabla \psi_1 \right)
= 0,
\\
\rho_0 \left( {\partial \psi_1 \over \partial t} + 
\nabla\psi_0 \cdot \nabla \psi_1 \right) = p_1,
\\
p_1 = c_s^2 \rho_1.
\end{eqnarray} 
Here we define $c_s^2 \equiv \partial p/\partial\rho$, the standard
definition for the speed of sound.

\Exercise Derive these equations. (The second equation, which comes
from the Euler equation, is a little tricky.)

These three first-order partial differential equations can be assembled
into one second-order partial differential equation which carries
exactly the same information: 
\begin{eqnarray}
\label{E:physical}
&&{\partial\over\partial t} 
\left( c_s^{-2} \rho_0 
\left( {\partial\psi_1\over\partial t} + {\vec v}_0 \cdot \nabla \psi_1
\right)
\right)
\nonumber\\
&&
\qquad
= \nabla \cdot \left( 
\rho_0 \nabla \psi_1 - c_s^{-2} \rho_0 {\vec v}_0
\left( 
{\partial \psi_1 \over \partial t} + {\vec v}_0 \cdot \nabla \psi_1 
\right)
\right).
\end{eqnarray}
This is a second-order partial differential equation for $\psi_1$ with
variable coefficients that depend only on the background field around
which we are linearizing. Once a solution $\psi_1(t,x)$ has been
obtained, the pressure fluctuations $p_1(t,x)$ and density fluctuations
$\rho_1$ follow directly from the linearized Euler equation and the
linearized equation of state.

\Exercise (Easy) Check this.

To turn this into a form suitable for obtaining a {\em spacetime}
interpretation, introduce four-dimensional coordinates via the usual
identification
\begin{equation}
x^\mu \equiv (t, \vec x).
\end{equation}
Now introduce a $4\times4$ matrix
\begin{equation}
g^{\mu\nu}(t,\vec x) \equiv 
{1\over \rho_0 c_s}
\left[ \matrix{-1&\vdots&-v_0^j\cr
               \cdots\cdots&\cdot&\cdots\cdots\cdots\cdots\cr
               -v_0^i&\vdots&(c_s^2 \delta^{ij} - v_0^i v_0^j )\cr } 
\right].               
\end{equation}
(Greek indices run from $0$--$3$, while Roman indices run from
$1$--$3$.) Define
\begin{equation}
g = \left[\det\left(g^{\mu\nu}\right)\right]^{-1}.
\end{equation}
Then the rather formidable-looking second-order partial
differential equation for $\psi_1$ can be very simply written as
\begin{equation}
\label{E:EOM}
{1\over\sqrt{-g}} \;
{\partial \over \partial x^\mu} \left( 
\sqrt{-g} \; g^{\mu\nu} \; {\partial \over \partial x^\nu} \psi_1 
\right) = 0.
\end{equation}

\Exercise (Straightforward) Check this. Calculate $g$. Show that
(\ref{E:EOM}) above is {\em identical} to (\ref{E:physical}).

%-------------------------------------------------------------------------
\subsection{Interpretation}
%-------------------------------------------------------------------------
%
Once you have reduced the equations of motion to the form
(\ref{E:EOM}), the last step is trivial: Just observe that this
equation is exactly that of a minimally coupled massless scalar field
propagating in a spacetime with inverse metric $g^{\mu\nu}(t,x)$. In
fact, the differential operator appearing in this equation is just the
d'Alambertian of the inverse metric $g^{\mu\nu}(t,x)$.

\Exercise Check that the matrix 
\begin{equation}
g_{\mu\nu}(t,\vec x) 
\equiv {\rho_0\over c_s} 
\left[ \matrix{-(c_s^2-v_0^2)&\vdots&-{(v_0)_j}\cr
               \cdots\cdots\cdots\cdots&\cdot&\cdots\cdots\cr
               -{(v_0)_i}&\vdots& \delta_{ij}\cr } \right].
\end{equation}
is the inverse of $g^{\mu\nu}$. Check that the signature of this
matrix is $(-,+,+,+)$. 

Thus we have demonstrated that the propagation of sound is governed by
an {\em acoustic metric} --- $g_{\mu\nu}(t,\vec x)$. This acoustic
metric describes a $(3+1)$--dimensional Lorentzian
(pseudo--Riemannian) geometry. The metric depends {\em algebraically}
on the density, velocity of flow, and local speed of sound in the
fluid. This is the essential difference between this acoustic
Lorentzian geometry and GR: The acoustic metric is governed by the
fluid equations of motion (continuity, Euler's equation, and the
barotropic equation of state) which constrain the background geometry,
and the Einstein equations of GR are not useful in this context. You
can certainly {\em calculate} the Ricci tensor and Einstein tensor for
this acoustic metric, but there is no justification for asking these
quantities to satisfy any particular constraint.

Note that although the underlying physics (fluid mechanics) is
completely non-relativistic, sharply separating the notions of space
and time, the fluctuations (sound waves) nevertheless couple to a
spacetime metric that places space and time in a unified framework.

\Exercise (Some tricky points.) Copy/extend the standard definitions
of black hole, event horizon, apparent horizon, and surface gravity
into this context. Check your ideas against the discussion below, and
the more detailed formulation in~\cite{Visser98a,Visser98b}.

%----------------------------------------------------------------------------
\section{Examples}
%-------------------------------------------------------------------------
\subsection{Nozzle}
%-------------------------------------------------------------------------
%
A particularly simple example of a non-trivial acoustic geometry is
provided by laminar fluid flow through a
nozzle~\cite{Unruh81,Visser98a}: As the nozzle narrows the fluid
speeds up. If the nozzle is sufficiently narrow the fluid velocity
will exceed the local speed of sound. (Doing this experimentally while
maintaining laminar flow is a rather difficult proposition.) If the
fluid velocity exceeds the speed of sound, then sound waves cannot
escape back out of the region of supersonic flow. Thus a region of
supersonic flow shares many of the properties normally associated with
a black hole (more properly, this region shares many of the properties
of the ergosphere of a black hole).

%-------------------------------------------------------------------------
\subsection{Vortex}
%-------------------------------------------------------------------------
%
A draining bathtub, with water swirling around the drain, is a useful
model for emphasizing the difference between an event horizon and an
ergosphere~\cite{Visser98a}. As one moves inwards towards the drain,
two interesting things happen: First, the magnitude of the fluid
velocity exceeds the speed of sound, and second (somewhat nearer the
drain) the {\em radial component} of the fluid velocity exceeds the
speed of sound.

The region in which the fluid velocity exceeds the speed of sound
defines the ergoregion. (It is impossible to stand still in the
ergoregion without producing a sonic boom.) Once the radial component
of the velocity exceeds the speed of sound, then acoustic disturbances
cannot escape from the region around the drain, and this defines the
event horizon.

If the motion is perfectly radial, (no swirling) the two notions
agree. It is only if there is a swirling motion near the drain that
the two notions need to be separated. This is the analog for fluid
dynamics of the GR behaviour of the metric near a rotating black hole:
For the Kerr metric the dragging of inertial frames implies that the
region in which one cannot remain at rest with respect to asymptotic
infinity [the ergoregion] is not the same as the region from which you
cannot escape to asymptotic infinity.

%-------------------------------------------------------------------------
\subsection{Supersonic Cavitation}
%-------------------------------------------------------------------------
%
It is experimentally very easy to set up a situation in which air
bubbles in water are induced to collapse at supersonic speeds. (Speeds
of up to Mach 4 are quite common.) Supersonic bubble collapse provides
an example of an apparent horizon (not an event horizon). It's an
apparent horizon because simply by waiting for bubble collapse to
stop, and the re-expansion phase to start, you can always be
guaranteed of getting a sound signal back out to spatial infinity ---
thus there cannot be a true event horizon (absolute horizon) in the
system.

Furthermore, this apparent horizon can exist for only a very short
time during each collapse cycle: By construction, the apparent horizon
lasts for less than one sound-crossing time.

Experiments of this type are normally set up to investigate the
phenomenon of {\em sonoluminescence}. Before anyone gets too carried
away, let me state explicitly that the visible light emitted in
sonoluminescence is not Hawking radiation associated with this
apparent horizon: (1) If anything, you should expect phonons, not
photons~\cite{Hochberg}; (2) The ``Hawking temperature'' estimated
from the acceleration of the bubble wall, while somewhat larger than
the most naive estimates based on Unruh's analysis~\cite{Unruh81}, is
still far too small to be relevant for
sonoluminescence~\cite{Hochberg}; (3) The fact that the apparent
horizon lasts for less than one sound-crossing time renders Hawking's
calculation moot. See below for more discussion of acoustic Hawking
radiation.

%-------------------------------------------------------------------------
\subsection{(Conformal) Schwarzschild flow}
%-------------------------------------------------------------------------
%
Can we find a fluid flow that exactly mimics the Schwarzschild
geometry? No, but we can get reasonably close: We can find a fluid
flow that has an acoustic metric that is conformal to that of
Schwarzschild spacetime.  Start by writing the Schwarzschild geometry
in \Painleve--Gullstrand form
\cite{Painleve,Gullstrand,Lemaitre,Kraus-Wilczek}
\begin{equation}
ds^2 = 
-\left(1-{2GM\over r}\right) \; dt^2
\pm \sqrt{2GM\over r} \; dr \; dt
+ dr^2 + r^2 \left( d\theta^2 + \sin^2\theta \; d\phi^2 \right).
\end{equation}
In this coordinate system the Schwarzschild geometry has been written
in such a way that {\em space} is flat, though {\em spacetime} is
curved.

\Exercise Find the coordinate transformation needed to go from any of
the more usual representations of Schwarzschild spacetime to this one.

\Exercise Demonstrate that for any spherically symmetric geometry (not
necessarily static, though you may want to consider that special case
first) it is always possible to find a coordinate system such that
{\em space} is flat. (So all the spacetime curvature can be forced
into the $g_{tt}$ and $g_{ti}$ components of the metric.)

\Exercise Take the general acoustic metric. Pick $c_s$ a
position-independent constant, $\vec v = \sqrt{2GM/r}\; \hat r$, and
$\rho \propto r^{-3/2}$. Check that this fluid flow satisfies the
equation of continuity~\cite{Visser98a}. Find the equation of
state. Find the background pressure distribution needed to satisfy the
Euler equation. Demonstrate that for this choice of fluid flow
\begin{equation}
(g_{\mu\nu})_{\mathrm{acoustic}} \propto 
r^{-3/2} (g_{\mu\nu})_{\mathrm{Schwarzschild}}.
\end{equation}
Finally, show that this is the best that can be done~\cite{Visser98a}.

%---------------------------------------------------------------------------
\section{Surface gravity}
%-------------------------------------------------------------------------
%
In the same way that one can define a surface gravity for a black hole
in General Relativity, it is also possible to set up a notion of
surface gravity for an acoustic black hole. Unruh
showed~\cite{Unruh81} that under certain conditions the surface
gravity is related to the normal derivative of the fluid velocity as
it crosses the event horizon, and is then equal to the physical
acceleration of the fluid as it crosses the event horizon.
\begin{equation}
g_H = c_s \; {\partial v\over \partial n} = a_{\mathrm{fluid}}.
\end{equation}
Unfortunately, this result is limited to the case where (1) the speed
of sound is position independent, and (2) the fluid crosses the event
horizon perpendicularly (which means the event horizon must be
identical to the ergoregion). The general result, derived
in~\cite{Visser98a}, is
\begin{equation}
g_H = 
{1\over2} \left|{\partial\over\partial n} (c_s^2 - v_\perp^2)\right| =
c_s\; \left|{\partial \over\partial n} (c_s - v_\perp) \right|.
\end{equation}
So a position dependent speed of sound, or a non-trivial ergoregion,
greatly complicates life.

\Exercise There are a large number of technical incantations required
to justify these formulae. Consider a null geodesic that just skims
the acoustic event horizon, and parameterize it by non-relativistic
Newtonian time. Show that this parameter is not an affine parameter
for the null geodesic, and show that the surface gravity measures the
extent to which Newtonian time fails to be an affine parameter. For
more details see~\cite{Visser98a}.

%---------------------------------------------------------------------------
\section{Acoustic Hawking radiation}
%-------------------------------------------------------------------------
%
With the build up we have seen so far, the discussion of acoustic
Hawking radiation~\cite{Unruh81,Visser93a,Visser98a,Comer} is almost
anticlimactic, and can be relegated to a series of exercises.

\Exercise Read the {\em original} paper demonstrating the existence of
Hawking radiation (the one in Nature~\cite{Hawking74}, compare
with~\cite{Hawking75}). Check that this derivation does not need or
use the Einstein equations. (The key feature of this original
derivation is that the black hole is quasi-static: there should be an
apparent horizon that lasts for a long time compared to the
light-crossing time for the black hole.)

\Exercise Compare this with some of the subsequent rederivations of
the Hawking radiation effect. For example, using analytic continuation
to Euclidean signature. Later derivations are technically slicker (and
more computationally efficient) but often obscure the underlying
physics.

\Exercise Verify that an acoustic black hole will emit a quasi-thermal
{\em phonon} spectrum with temperature
\begin{equation}
k T_H = {\hbar g_H\over2\pi c_s} = 
{\hbar\over2\pi} \left| {\partial\over\partial n} (c_s - v_\perp) \right|.
\end{equation}
Near the horizon, the spectrum is almost exactly thermal. As the
phonons move away from the event horizon they are to some extent
back-scattered by the acoustic metric. Exactly the same phenomenon
occurs in GR and is the origin of the famous grey-body factors ---
even for a Schwarzschild black hole the emission spectrum is not
exactly Planckian.

\Exercise Put in some numbers. Verify that
\begin{equation}
T_H = 1.2 \times 10^{-6} \hbox{ K mm} 
\left[ {c_s\over 1 \hbox{ km/sec} } \right]
\left[ {1\over c_s} {\partial (c_s - v_\perp) \over \partial n} \right].
\end{equation}
Thus for supersonic flow of water through a 1 mm nozzle, $T_H \approx
10^{-6} \hbox{ K}$. If this number was just a little bit better, we
could reasonably hope to build laboratory experiments to verify this
acoustic Hawking effect. Temperatures of $10^{-6}$ Kelvin are not by
themselves completely out of reach, though you would certainly not be
using water as the working fluid. The real issue is that of detecting
a thermal phonon spectrum at this temperature, while maintaining
laminar supersonic flow for the working fluid.

\Exercise Verify that the existence of Hawking radiation is a purely
kinematic effect. Hawking radiation will occur in any Lorentzian
geometry that contains an event horizon, independent of what the
dynamical equations underlying the geometry are.  See for
instance~\cite{Brout,Jacobson95,Jacobson96,Corley-Jacobson96,Corley-Jacobson97,Corley97a,Corley97b,Reznik96,Reznik97,Volovik,Volovik-Vachaspati}.

%---------------------------------------------------------------------------
\section{Horizon entropy}
%-------------------------------------------------------------------------
%
In contrast to Hawking radiation, which is a purely kinematic effect,
the notion of black hole entropy is intimately tied up with the
dynamics of the geometry. In fact, if the dynamics of the geometry is
governed by a Lagrangian that depends only on the metric (and thus
implicitly on the Riemann tensor of the metric) then there is a rather
general formula for black hole entropy~\cite{Visser93b}
\begin{equation}
S = {k\over\ell_P^2} 
\int {\delta {\cal L} \over \delta R_{abcd} } 
\; \epsilon_{ab} \; \epsilon_{cd} \; \sqrt{ {}^{(2)} g} \; d^2 x.
\end{equation}
Here the integral runs over the two-dimensional event horizon, and
$\epsilon$ denotes the two-dimensional Levi-Civita symbol defined on
the event horizon. In particular, suppose (for simplicity) that we have
\begin{equation}
{\cal L} = {1\over8\pi G} \; R \left( 1 + \sum_{n=1}^\infty a_n R^n \right),
\end{equation}
then for a spherically symmetric black hole
\begin{equation}
S = {1\over4} \; {k \over\ell_p^2} \; A \; 
\left( 1 + \sum_{n=1}^\infty b_n A^{-n} \right).
\end{equation}
This example is enough to drive home a key point: Entropy equals area
(plus corrections) {\em if and only if} the Lagrangian is
Einstein--Hilbert (plus corrections)~\cite{Visser93b,Visser93c}. So
calculations of black hole entropy are implicitly calculations of the
Lagrangian governing the geometry~\cite{Visser98b}.

There are some interesting quirks of history here: Historically
Bekenstein's notion of black hole entropy came first, and Hawking
radiation was discovered as a side effect of trying to make the notion
of black hole entropy consistent with ordinary thermodynamics. But now
we see that Hawking radiation is a much more primitive concept, one
that is more fundamental than the black hole entropy it helped
explain. In fact Hawking radiation makes perfectly good sense even in
situations in which the notion of black hole entropy is entirely
meaningless~\cite{Visser98b}.

This also has implications for string theory~\cite{Visser98b}: We have
known since the mid 1980's that the low-energy (sub-Planckian) limit
of essentially any string theory is a theory of curved spacetime with
dynamical equations derived from an action that is the
Einstein--Hilbert action (plus corrections). Thus in {\em any}
phenomenologically interesting string theory black holes must have an
entropy that is proportional to the area (plus corrections). In a
certain sense, complicated state-counting calculations in the
underlying string theory can be viewed as consistency checks that
verify that the low-energy dynamics is what you thought it was. There
are suspicions, though not a complete proof, that it might be able to
formalize this statement by rephrasing the state-counting calculations
directly in terms of the low-energy degrees of freedom. See for
instance the Horowitz--Polchinski ``Correspondence
Principle''~\cite{Correspondence,Correspondence2} or
Carlip's~\cite{Carlip} and Solodukhin's~\cite{Solodukhin} analysis in
terms of a the central charge of an appropriate Virasoro algebra
attached to the event horizon.

%---------------------------------------------------------------------------
\section{Discussion}
%-------------------------------------------------------------------------
%
The acoustic model for Lorentzian spacetime is a very good toy model
for forcing you to think long an hard about fundamental issues in GR
(and fluid mechanics, and even String Theory). It forces you to
sharply separate those aspects of black hole physics that are purely
kinematical from those parts that are intrinsically dynamical. It
forces you to think about the universality of Lorentzian geometry:
Even completely non-relativistic fluid dynamics has a Lorentzian
spacetime hiding inside it. It allows you to formulate in a coherent
manner possible approaches to the breakdown of Lorentz invariance
(though I have not said anything about this topic in this talk).

A key result that I would like the reader to appreciate is this:
Hawking radiation is kinematics; Black hole entropy is dynamics.

Finally, an observation: It is often quite remarkable how much really
deep and fundamental physics can be found hiding in quite unexpected
places.

%--------------------------------------------------------------------------

%--------------------------------------------------------------------------

%INDEX%%%%%%%%%%%%%%%%%%%%%%%%%%%%%%%%%%%%%%%%%%%%%%%%%%%%%%%%%%%%%%%
\clearpage
\addcontentsline{toc}{section}{Index}
\flushbottom
\printindex
%%%%%%%%%%%%%%%%%%%%%%%%%%%%%%%%%%%%%%%%%%%%%%%%%%%%%%%%%%%%%%%%%%%%%

%--------------------------------------------------------------------------
\end{document}